\documentclass[12pt]{iopart}
\usepackage{graphicx}

\begin{document}

\title{ Diffusion Anomaly in an  Associating Lattice Gas Model }

\author{Marcia M. Szortyka}

\address{Instituto de F\'{\i}sica, Universidade Federal do Rio Grande do
Sul, Caixa Postal 15051, 91501-970, Porto Alegre, RS, Brazil}

\author{Marcia C. Barbosa\footnote[3]{To
whom correspondence should be addressed (marcia.barbosa@ufrgs.br)}}

\address{Instituto de F\'isica, Universidade Federal do Rio Grande do
Sul, Caixa Postal 15051, 91501-970, Porto Alegre, RS, Brazil}

\date{\today}
\begin{abstract}

\end{abstract}

We investigate the relation
between thermodynamic and dynamic properties
of an associating lattice gas (ALG) model. The ALG combines a two dimensional 
lattice
gas with particles interacting through a soft core potential and orientational
degrees of freedom. From the competition between the directional 
attractive forces and the soft core
potential results two liquid phases, double criticality and density anomaly. 
We study the mobility of the molecules in this model by
calculating    the diffusion constant at a constant temperature, $D$. 
We show that $D$
has a maximum at a density $\rho_{max}$ and a minimum at a density
$\rho_{min}<\rho_{max}$. Between these densities the diffusivity
differs from the one expected for normal liquids. We also show that in
the pressure-temperature phase-diagram
the line of extrema in diffusivity is close to the liquid-liquid critical point and it 
is inside the temperature of maximum density (TMD) line.

\maketitle


\section{\label{sec1}Introduction}


Water is anomalous substance in many respects.
Most liquids contract upon cooling. This is not the case of water, a liquid
where the specific volume at ambient pressure starts to increase 
when cooled below $T=4 ^oC$ \cite{Wa64}. Besides, in a certain
range of pressures, also exhibits an anomalous increase of compressibility 
and specific heat upon cooling \cite{Pr87}-\cite{Ha84}.

Far less 
known are its dynamics anomalies: while for most
materials diffusivity decreases with increasing pressure,
liquid water has an opposite behavior in a large region
of the phase diagram 
\cite{St99}-\cite{Ne02}.
The increase of diffusivity of water as the pressure
is increased is related to the competition between the local
ordered tetrahedral structure of the first neighbours and the distortions
of the structure of the first and second neighbours. In the region
of the phase diagram where this ordered structure is dominant,
increasing pressure implies breaking first neighbours
hydrogen bonds what allow for interstitial second neighbours to be in a closer
approach. The interactions are thus weakened and therefore,
although the  system  is more dense, it has a larger
mobility. In this sense, a good model for water and tetrahedral 
liquids should not only
exhibit thermodynamic but also dynamic anomalies.
In SPC/E water, the region of the pressure-temperature (p -T) 
phase diagram where the density anomaly appears is contained 
within the region of the p -T phase diagram where anomalies in 
the diffusivity are present  \cite{Er01,Ne01}.

It was proposed a few years ago
that these anomalies are related to a second critical
point between two liquid phases, a low density liquid
(LDL) and a high density liquid (HDL) \cite{Po92}. This critical point 
was discovered
by computer simulations. This work   suggests that this
critical point is located at the 
supercooled region beyond  the line of
homogeneous nucleation and thus cannot be experimentally measured. 
Even if this limitation,  this hypothesis has been supported by indirect experimental 
results \cite{Mi98,angell}.

Water, however, is not 
an isolated case. There are also other examples of  tetrahedrally
bonded molecular liquids such as phosphorus \cite{Ka00,Mo03}
and amorphous  silica \cite{La00} that also are good candidates for having
two liquid phases. Moreover, other materials such as liquid metals
\cite{Cu81} and graphite \cite{To97} also exhibit thermodynamic anomalies.
Unfortunately, a closed theory 
giving the relation between the form of the interaction potential 
and the presence of the anomalies is still missing.

It was observed that water has both diffusion and density anomalous
behavior and that they are located in the same region 
of the $p$-$T$ phase diagram.
It is reasonable to think that the dynamics and thermodynamics
  are deeply related.
In this case establishing the connection between the anomalous 
behavior of the diffusion constant and the density anomaly
  is a fundamental step towards understanding
the source of the anomalies. In order to test this assumption,
the dynamics of a number of models in which density anomaly is present
were studied \cite{Ne06,Ol06}.  Netz et al \cite{Ne06} studied
a system of
molecules interacting by a 
purely repulsive ramp-like discretized potential consisting of 
a $n$ number of steps of equal size.  If $n$ is small, 
the region on the $p$-$T$ phase diagram in which
the density anomaly is present  
encloses the region in which the  diffusion anomaly exists. If $n$ is
large, the potential becomes equivalent to a smooth ramp \cite{Ol06} and 
 the region $p$-$T$ phase diagram in which
the diffusion anomaly is present  
encloses the region in which the  density anomaly exists.
This behavior resembles the one expected for water.
Unfortunately these two models do not exhibit a second 
critical point (or even a first critical point) because
the interaction potential in both cases is purely repulsive.
Therefore no connection with criticality and dynamic
and thermodynamic anomalies could be made.

The difficulty in including the attractive interactions
in these approaches is that within continuous potentials attractive terms
usually move the density anomaly to the metastable region of the 
 $p$-$T$ phase diagram. Moreover 
 continuous potentials usually lead to crystallization 
at the region where the critical point would be expected. The 
analysis of the presence of both the two liquid phases and the
critical point becomes indirect.
 
In order to  circumvent these difficulties, our model system is 
the a lattice gas with  ice
variables \cite{Be33} which allows for a low density ordered structure \cite{He05a,He05b}.
 Competition between the filling
up of the lattice and the formation of an open four-bonded orientational
structure
 is naturally introduced in terms of the ice bonding variables and
no \emph{ad
 hoc} introduction of density or bond strength variations is
needed. Our approach
 bares some resemblance to that of some
continuous models\cite{Si98,Tr99,Tr02},
which, however, lack
entropy related to hydrogen distribution on bonds. Also, the reduction of
phase-space imposed by the lattice allows construction
 of the full phase
diagram from simulations, not always possible for continuous
 models
\cite{Si98}. The associating lattice gas 
 model (ALG)\cite{He05a,He05b} is
able to exhibit for a convenient set of parameters both
 density anomalies and
the two liquid phases. In the framework of the ALG we address two
questions: (i) is the presence of 
diffusion anomaly
related to the presence of density anomaly? (ii) if so, what
is the hierarchy between the two anomalies and the presence of a 
second critical point?
 We show that the two anomalies are located in the same region
of the $p$-$T$   phase diagram, close to the second critical
point and that  the region on the $p$-$T$ phase diagram in which
the density anomaly is present  
encloses the region in which the  diffusion anomaly exists.


\section{\label{sec2}The Model}

We consider the ALG that is 
 a two dimensional lattice gas model on a triangular lattice as 
introduced by \emph{Henriques} and \emph{Barbosa} 
\cite{He05a,He05b}. The sites that can be empty or occupied.  Besides 
the occupational variable, $\sigma_i$, which assumes the value 
$\sigma_{i}=0$ if the site is empty, or $\sigma_{i}=1$ if the site is full, 
there are  other  six bond variables, $\tau_{i}^{ij}$, representing 
the possibility of hydrogen bond formation
 between the neighbours sites. Four bonding variables are the 
ice bonding arms, two donor with $\tau_{i}^{ij}=1$ and two acceptor with 
$\tau_{i}^{ij}=-1$. 
The other two arms are taken as 
inert, with $\tau_{i}^{ij}=0$, and are always opposite to each other, 
as illustrated in Fig. 1. Because 
there are no restrictions on the bonding arms positions, the particles 
are allowed to be in 
eighteen different states.

The Hamiltonian of the model has two kinds of interactions: one isotropic 
attractive, like van der Waals, and 
a repulsive orientational hydrogen bonding. The energy of the system is 
given by
\begin{equation}
\label{E}
E=\left(-v+2u\right)\sum_{\left\langle i,j\right\rangle }\sigma_{i}
\sigma_{j}+u\sum_{\left\langle i,j\right\rangle }\sigma_{i}\sigma_{j}
\sum_{k=1}^{6}\sum_{l}^{*}\left[\left(1-\tau_{i}^{k}\tau_{j}^{l}
\right)\tau_{i}^{k}\tau_{j}^{l}\right]\;\;,
\end{equation}
where \emph{-v+2u} is the 
van der Waals interaction, \emph{-2u} is 
the hydrogen bonds interaction, \( \sigma _{i}=0,1 \) are 
occupation variables, \( \tau _{i}^{k}=0,\pm 1 \) represent the 
bonding arms variables as illustrated in Fig.~1, the summation over \emph{k} is
over the arms of the site we are considering and the summation over \emph{l} is
over the six neighbours arms that especific point to the site arm. Two neighbour 
sites $i$ and $j$ form an hydrogen bond if the product
between their pointing arms is equal to $-1$, in other words, we need $\tau^{k}_i \tau^{l}_{j}=-1$. 
In spite of each molecule has six neighbours, only four hydrogen bonds are 
allowed for site. For each pair of occupied sites that 
form a hydrogen bond, an energy \emph{-v} is attributed, while 
for non-bonding pairs of occupied sites, the energy is \emph{-v+2u}. 

\begin{figure}
\begin{center}
\includegraphics[scale=0.8,width=10cm]{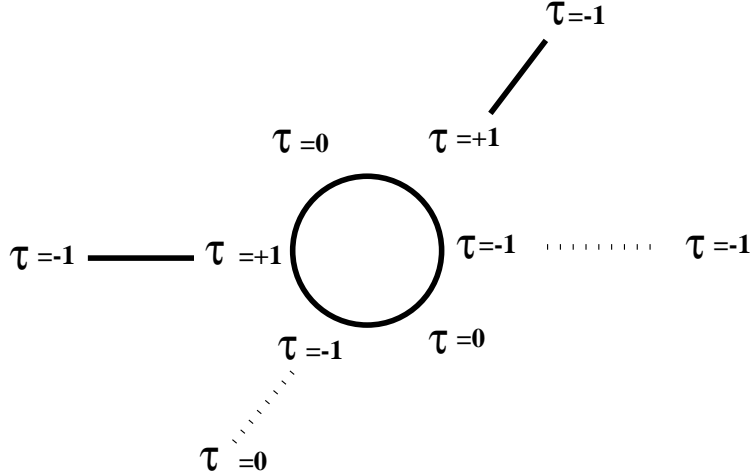}
\end{center} 
\caption{Illustration of the model. Each site $i$ can be 
empty or occupied and has  six variables, $\tau_i^k$, one
for each arm. If $\tau_i^k=0$ no bond is formed in spite of the
configuration of the arm of the neighbour site, if $\tau_i^l=\pm 1$ and the
arm of the neighbour site has $\tau_j^l=\mp 1$, a bond is formed.}
\label{fig1}
\end{figure}

\begin{figure}
\begin{center}
\includegraphics[clip=true,scale=0.8,width=10cm]{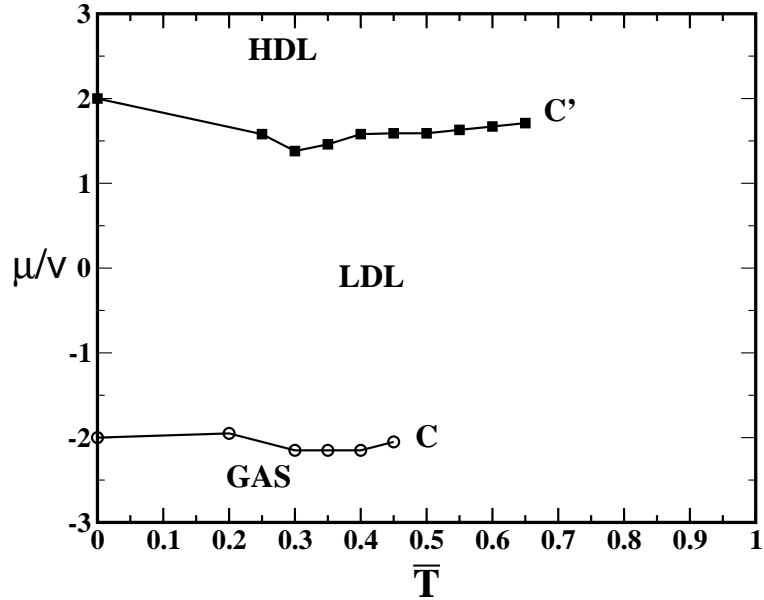}
\end{center} 
\caption{ Phase diagram showing reduced chemical vs. reduced temperature for 
$u/v=1$. The 
filled symbols  represent  the  LDL-HDL  coexistence line. The empty symbols
indicate the gas-LDL coexistence line. The coexistence 
at  zero temperature  at  $\overline{\mu}=-2$  and $\overline{\mu}=2 $ are exact.
}
\label{fig2}
\end{figure}

\begin{figure}
\begin{center}
\includegraphics[clip=true, scale=0.8,width=10cm]{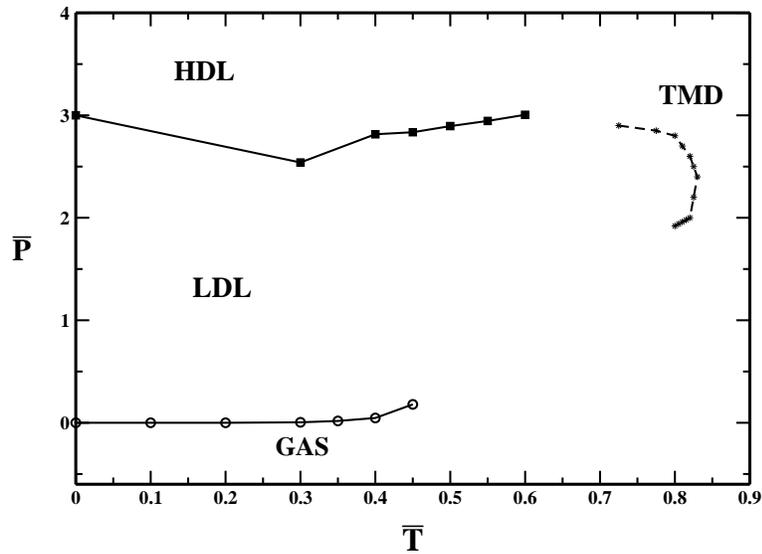}
\end{center} 
\caption{ Phase diagram showing reduced pressure vs. reduced temperature for 
$u/v=1$. The 
filled symbols  represent  the  LDL-HDL  coexistence line. The empty symbols
indicate the gas-LDL coexistence line. The coexistence 
at  zero temperature  at  $p/v=3$  and $p/v=0$ are exact. The dashed line is
the $TMD$. 
}
\label{fig3}
\end{figure}


 The phase diagram of the 
reduced chemical potential, $\overline{\mu}=\mu/v$, versus
reduced temperature, $\overline{T}=k_BT/v$  reproduced in 
Fig. 2 was
originally obtained by Henriques and Barbosa  \cite{He05a} using Monte Carlo
simulations for a triangular lattice with $L^2=100$ lattice sites.
At low 
temperatures and  chemical potentials, the system 
is in the gas phase. As the 
the chemical  is increased at constant 
temperature, there is a first-order 
phase transition between the gas to a low density liquid 
phase $(LDL)$. As the chemical potential  is increased even
further, there is another first-order transition between the LDL phase 
a high density liquid phase $(HDL)$. The two transition end at critical 
points. 

The reduced pressure, $\overline{P}=P/v$
 versus reduced temperature phase diagram 
was also obtained by Henriques and Barbosa  \cite{He05a}
and it is illustrated in Fig.~3. In 
the $\overline{P}-\overline{T}$ plane, close 
to the   $LDL-HDL$ critical point, there is a region 
of temperature of maximum density $(TMD)$
 for a given pressure.
 Surprisingly
 the location of the continuous transitions, the critical points and 
the density anomaly are not very sensitive to the systems size. 
 Some test runs  were done for $L^2=400,2500,4900$. The difference
between the  critical points obtained by applying finite 
size scaling to those runs
and the result for $L^ 2=100$ is quite small. So, 
for simplicity all the detailed
study of the model properties and the full phase diagrams was undertaken for
an $L^2=100$ lattice sites. 

This simple model exhibits two liquid phases, two critical points
and density anomaly similar to the ones present in the SPC/E model 
for water and in other potential models for water. In the next
section we will investigate if it also has diffusion anomaly 
and how this anomaly is related to the $TMD$.

\section{\label{sec3}Diffusion}

For studying the mobility, we have performed
 Monte Carlo simulations of a system
of $n$ particles interacting as 
specified by  the Hamiltonian
Eq.(\ref{E}) in  a triangular  lattice with $L^2=100$ lattice sites. 
 The procedure for computing the diffusion coefficient
goes as follows. The system is equilibrated 
at a  fixed chemical potential and temperature.  In equilibrium
this system has $n$ particles.  
Starting from this equilibrium configuration at a time $t=0$, each
one of these  $n$
particles is allowed to move to an empty neighbour site
randomly chosen. The move
is accepted  if the total energy of the system 
is reduced by the move, otherwise it is accepted with
a probability $\exp(\Delta E /k_BT)$ where $\Delta E$ is 
the difference between the energy of the system after and before the
move. After repeating this procedure  $nt$ times, the mean 
square displacement per particle  at a time $t$
is computed by
\begin{equation}
\langle \Delta r(t)^2 \rangle =\langle\left(\textbf{r}(t)-
\textbf{r}(0)\right)^2 \rangle\;\;,
\label{D}
\end{equation}
where  $\textbf{r}(0)$ is the particle
position at the initial time and  $\textbf{r}(t)$ is the 
particle position at a time $t$. In Eq.~(\ref{D}), the average 
is taken over all particles and over 
different initial configurations.
The diffusion coefficient is then obtained from the relation
\begin{equation}
D=\lim_{t\rightarrow \infty}\frac{\langle \Delta r(t)^2 \rangle}{4t}\; .
\end{equation}
Since the time is measured in Monte Carlo time steps  and 
the distance in number of lattice distance, a dimensionless 
diffusion coefficient is defined as
\begin{equation}
\overline{D}=\lim_{t\rightarrow \infty}\frac{\langle \Delta
\overline{r}(t)^2 \rangle}{4\overline{t}}\; .
\label{D}
\end{equation}
where $\overline{r}=r/a$ and $a$ is the distance 
between two neighbour sites and 
$\overline{t}=t/t_{MC}$
is the time in Monte Carlo steps. 

 In order to find
if our system
exhibits diffusion anomalies, we
have analysed how $\overline{D}$ varies with 
the number density $\rho=n/L^2$ for a fixed temperature.
 For each temperature and chemical potential, 200 samples were 
obtained and an average over samples
was made.

\begin{figure}
\begin{center}
\includegraphics[clip=true,scale=0.8,width=10cm]{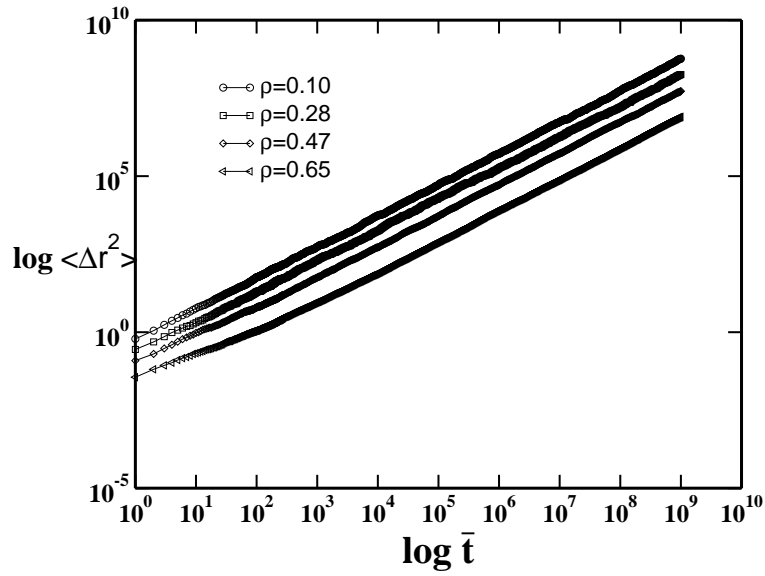}
\end{center} 
\caption{Mean square displacement vs. reduced 
time in logarithm scale for $\overline{T}=0.8$ and densities
from top to bottom $\rho= 0.1,0.28,0.47,0.65$. }
\label{fig4}
\end{figure}

\begin{figure}
\begin{center}
\includegraphics[clip=true,scale=0.8,width=10cm]{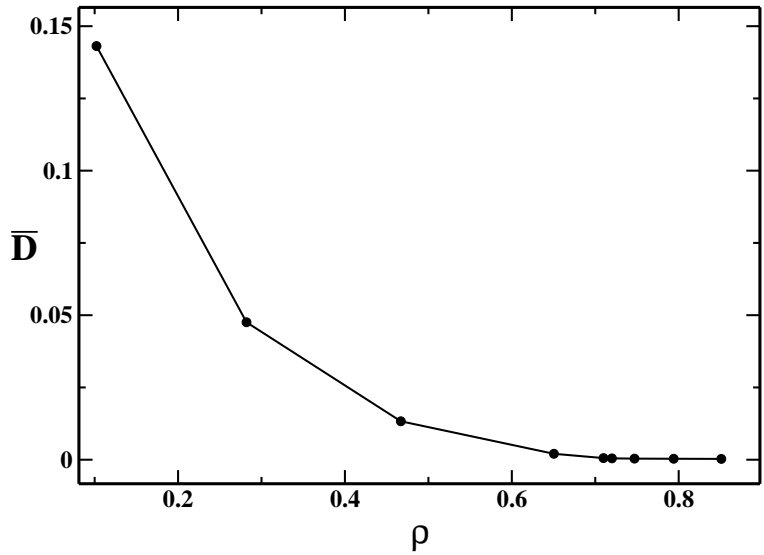}
\end{center} 
\caption{Reduced Diffusion coefficient vs. density for $\overline{T}=0.8$.
$\overline{D}$ increases with the reductions of the density as in a normal
liquid.  }
\label{fig5}
\end{figure}


\begin{figure}
\begin{center}
\includegraphics[clip=true,scale=0.8,width=10cm]{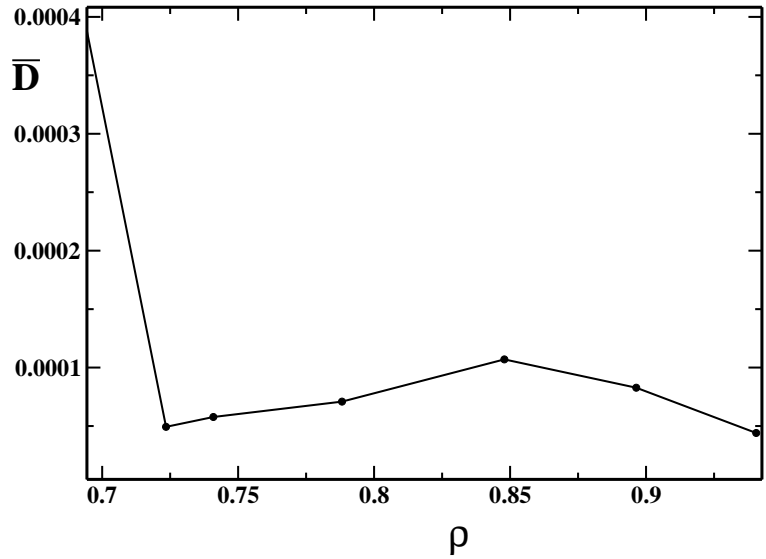}
\end{center} 
\caption{Reduced Diffusion coefficient vs. density for $\overline{T}=0.75$.}
\label{fig6}
\end{figure}


\begin{figure}
\begin{center}
\includegraphics[clip=true,scale=0.8,width=10cm]{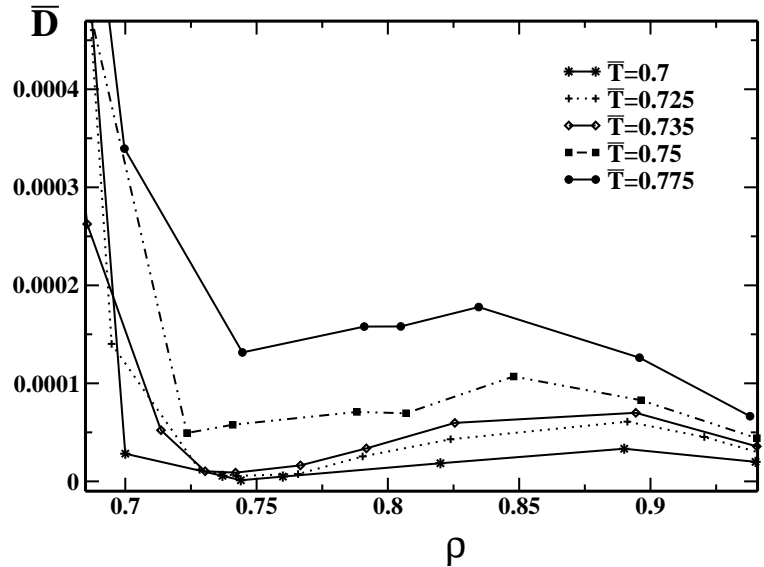}
\end{center} 
\caption{Reduced Diffusion coefficient vs. density for $T=0.7,0.725,0.735,0.75,0.775$.}
\label{fig7}
\end{figure}


\begin{figure}
\begin{center}
\includegraphics[clip=true,scale=0.8,width=10cm]{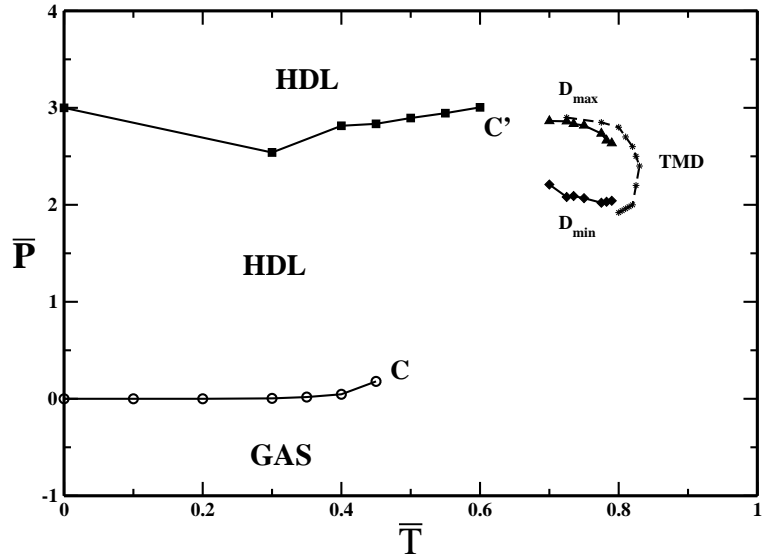}
\end{center} 
\caption{Reduced pressure vs. reduced temperature phase diagram showing the 
the two liquid phase, two critical points, the density anomaly and the diffusion anomaly regions.}
\label{fig8}
\end{figure}


\begin{figure}
\begin{center}
\includegraphics[clip=true,scale=0.8,width=10cm]{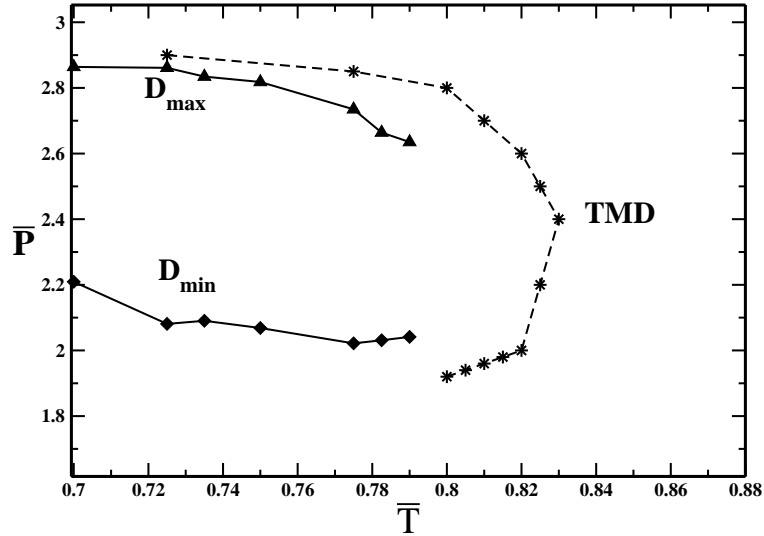}
\end{center} 
\caption{Zoom of the reduced pressure vs. reduced temperature phase diagram showing the density anomaly and diffusion anomaly regions.}
\label{fig9}
\end{figure}

At low chemical potentials and $\overline{T}=0.8$, $\rho$ is small and 
the mean square displacement  has only the
diffusive regime as illustrated in Fig.~4. As the chemical potential
is increased, $\rho$ also increases and 
the ballistic regime appears. 
Using the ballistic regime, the mean square displacement was then 
computed for densities 
ranging from $\rho=0.1$ to $\rho=0.9$ ( not illustrated in Fig.~4 for
clarity).
From this data together with Eq.~(\ref{D}), we have obtained
the reduced diffusion coefficient, $\overline{D}$,  versus 
density, $\rho$, shown in 
Fig.~5. In this case, $\overline{D}$
increases with the decrease of the density
what represents the behavior expected for
normal liquids. 
Simulations 
for higher temperatures are consistent with this result.

However, this is not the case at lower temperatures.
For instance, the   mean square displacement  
for  $\overline{T}=0.75$ exhibits a peculiar behavior
different from  the one observed in Fig.~4.
At the diffusive regime, for a certain density range,
the slope of $\langle(\Delta r)^2\rangle$ instead of increasing with
the decrease of density, decreases with it. Consequently
  the isochores 
cross each other.
The  reduced diffusion coefficient, $\overline{D}$ ( shown in Fig.~6) 
at very low densities decreases as $\rho$ increases like in 
a normal liquid.
However, as the density is increased $\overline{D}$ has a minimum 
at  $\rho_{Dmin}$, and  increases with
the increase of density from $\rho_{D_{min}}<\rho<\rho_{D_{max}}$.
Increasing the density above $\rho_{Dmax}$,  $\overline{D}$ decreases
as in a normal liquid. Therefore, there is a region
of densities $\rho_{Dmax}>\rho>\rho_{Dmin}$ where
the diffusion coefficient is anomalous, increasing
with density. This behavior is similar to the 
diffusion anomaly present in SPC/E water
and ramp and shoulder models. 
A diffusion anomaly in the ALG model  is 
observed in the range of temperatures $0.8>\overline{T}>0.6$
illustrated in Fig.~7.

The region in the $p$ -$T$ plane where there is an anomalous behavior in
 the diffusion is bounded by $(T_{D{\rm min}},P_{D{\rm min}})$ and
 $(T_{D{\rm max}},P_{D{\rm max}})$ and their location is shown
in Fig.~8.
The region of diffusion anomalies $(T_{D{\rm
 max}},P_{D{\rm max}})$ and $(T_{D{\rm min}},P_{D{\rm min}})$ 
lies inside the region of density anomalies ( for detail see fig.~9)
what differs from the behavior observed  in SPC/E water \cite{Ne01}
but coincides with the behavior shown for non smooth ramp-like
potentials \cite{Ne06}.

\section{\label{sec4}Conclusions} 

In this paper we have addressed two questions: (i) is the presence of 
diffusion anomaly
related to the presence of density anomaly? (ii) if so, what
is the hierarchy between the two anomalies and the presence of a 
second critical point?

For tracking the answer of these two questions
we have investigated the behavior of the diffusion coefficient
 in  the associating
 lattice gas model. This
simple model is suitable for addressing  our two questions since
it  exhibits two liquid phases and a 
line of density anomalies (TMD) \cite{He05a,He05b}.

Using Monte Carlo simulations we computed the 
 the mean square displacement with time. From the 
slope of this curve at the diffusive regime, the 
diffusion coefficient was derived. This procedure
was implement for different temperatures and densities.
We found that for high temperatures, particles in our
model system diffuse as a normal liquid. At
 low temperatures, however,  the
diffusion coefficient  has an interval
of densities $\rho_{D_{min}}<\rho<\rho_{D_{max}}$
 where $\overline{D}$
increases with the increase of density. 
Therefore, the presence of a density anomaly
 seems to be associated with the 
presence of diffusion anomaly, confirming observations
made in other models \cite{Ol06,Ku05} and in water 
\cite{Ne01,Ne02b,Ne02}. This seems to indicate
that as the particles gain more energy by being 
close together, this gain facilitates the mobility.

For addressing the second question, we have located
in the $p$-$T$ phase diagram the 
pressure of the density of maximum diffusivity and 
the pressure of the density of minimum diffusivity.
Since for each temperature there is one $\rho_{D_{min}}$
and there is  one  $\rho_{D_{max}}$, the diffusivity extrema
is composed of two lines at the $p$-$T$ phase-diagram. 
By comparing the $TMD$ with the lines of 
diffusivity extrema, we found that the region in the $p$-$T$ plane of
diffusion anomaly  is located inside
the region of the density anomaly and that both
are in the vicinity of the second critical point.
This suggests that in models where anomalies are present, the 
second critical point might arises if a attractive term  in
the potential would be included.  
The hierarchy between the anomalies resembles the one observed in the
purely repulsive ramp-like discretized potential \cite{Ne06}. The link
between the two models is the presence of two competing
interaction distances and the non smooth transition between them. 
The first ingredient seems to be the one that defines the presence
of the anomalies, while the second might govern the hierarchy between
them \cite{Ne06}\cite{Ol06}\cite{Ku05}. Similar behavior
should be expected in other models where the density anomaly is 
also present \cite{Fr03}-\cite{Ca05}

\subsection*{Acknowledgements}

We thank to Jeferson Arenzon for having helped us with the 
computational method of diffusion in Monte Carlo. We thank
and the Brazilian science agencies CNPq, Capes, Finep and Fapergs 
for financial support.

\bigskip

\bigskip



\begin{thebibliography}{10}


\bibitem{Wa64} R. Waller, Essays of Natural Experiments, Johnson Reprint corporation, New York , 1964.

\bibitem{Pr87} F. X. Prielmeir, E. W. Lang, Rl. Speedy, H. -D.  L\"udemann, Phys. Rev. Lett. 59, 1128 (1987).

\bibitem{Pr88} F. X. Prielmeier, E. W. Lang, R. J. Speedy, H. -D. L\"udemann, B. Bunsenges, Phys. Chem. 92, 1111 (1988).

\bibitem{Ha84} L. Haar, J. S. Gallagher, G. S. Kell, NBS/NRC Steam Tables. Thermodynamic
and Transport Properties and Computer Programs for Vapor and Liquid States of Water in SI Units, Hemisphere Publishing Co., Washington DC, 1984, pp271-276.


\bibitem{St99}F. W. Starr, F. Sciortino and H. E. Stanley, Phys. Rev. E
{\bf 60}, 6757 (1999); F. W. Starr, S. T. Harrington, F. Sciortino and
H. E. Stanley, Phys. Rev. Lett. {\bf 82}, 3629 (1999).

\bibitem{Ga96}P. Gallo, F. Sciortino, P. Tartaglia and
S. -H. Chen, Phys. Rev. Lett. {\bf 76}, 2730 (1996);
F. Sciortino, P. Gallo, P. Tartaglia and S. -H. Chen,
Phys. Rev. E {\bf 54}, 6331 (1996);
S. -H. Chen, F. Sciortino and P. Tartaglia, \emph{ibid.}
{\bf 56}, 4231 (1997); F. Sciortino, L. Fabbian, S. -H. Chen
and P. Tartaglia \emph{ibid.} {\bf 56}, 5397 (1997).

\bibitem{Ha97}S. Harrington, P. H. Poole, F. Sciortino and
H. E. Stanley, J. Chem. Phys. {\bf 107}, 7443 (1997).

\bibitem{Sc91}F. Sciortino, A. Geiger and H. E. Stanley, Nature (London)
{\bf 354}, 218 (1991); J. Chem Phys. {\bf 96}, 3857 (1992).

\bibitem{Er01}J. R. Errington and P. G. Debenedetti, Nature (London)
{\bf 409}, 318 (2001). 

\bibitem{Ne01}P. A. Netz, F. W. Starr, H. E.  Stanley and
M. C. Barbosa, J. Chem. Phys. {\bf 115}, 344 (2001).

\bibitem{Ne02a} P. A. Netz, F. W. Starr, M. C. Barbosa and
H. E. Stanley, Physica A {\bf 314}, 470 (2002).

\bibitem{Ne02b} H. E. Stanley, M. C. Barbosa, S. Mossa, P. A. Netz,
F. Sciortino, F. W. Starr and M. Yamada, Physica A {\bf 315}, 281
(2002).

\bibitem{Ne02} P. A. Netz, F.W. Starr, M. C. Barbosa and 
H. E. Stanley, J. Mol. Liquids {\bf 101}, 159 (2002).


\bibitem{Po92}P.H. Poole, F. Sciortino, U. Essmann, and H.~E. Stanley, Nature \t
extbf{360},
324 (1992); Phys. Rev. E \textbf{48}, 3799 (1993); F. Sciortino, P.H. Poole,
U. Essmann, and H.E. Stanley, Ibid \textbf{55}, 727 (1997); S. Harrington, R.
Zhang, P.H. Poole, F. Sciortino, and H.E. Stanley, Phys. Rev. Lett. \textbf{78},
2409 (1997). 

\bibitem{Mi98} O. Mishima, H. E. Stanley, Nature 396, 329 (1998).

\bibitem{angell}R.J. Speedy and C.A. Angell, J Chem Phys 65, 851 (1976).


\bibitem{Ka00}Y. Katayama, T. Mizutani, W. Utsumi, O. Shimomura, 
M. Yamakata and K. Funakoshi, Nature {\bf 403}, 170 (2000). 

\bibitem{Mo03}G. Monaco, S. Falconi, W.A. Crichton and M. 
Mezouar, Phys. Rev. Lett. {\bf 90}, 255701 (2003).

\bibitem{La00}D.J. Lacks, Phys. Rev. Lett. {\bf 84}, 4629 (2000). 

\bibitem{Cu81}P. T. Cummings and G. Stell, Mol. Phys. {\bf 43}, 1267 (1981).

\bibitem{To97}M. Togaya, Phys. Rev. Lett. {\bf 79}, 2474 (1997).


\bibitem{Ne06}P. A. Netz, S. Buldyrev, M. C. Barbosa and H. E. Stanley
Phys. Rev. E {\bf 73}, 061504 (2006).

\bibitem{Ol06}A. B. de Oliveira, P. A. Netz, T. Colla, and M. C. Barbosa,
J. Chem. Phys. {\bf 124}, 084505 (2006). 


\bibitem{Be33}J.D. Bernal and R.H. Fowler, J. Chem. Phys. 1, 515 (1933).



\bibitem{He05a} Vera B. Henriques and Marcia C. Barbosa, Phys. Rev. E {\bf 71} (2005).

\bibitem{He05b}Vera B. Henriques, Nara Guisoni, Marco Aurelio Barbosa, Marcelo
Thielo and Marcia C. Barbosa,  Molecular Physics  {\bf 103}, 3001 (2005).


\bibitem{Si98}K.A.T. Silverstein, A.D.J.Haymet and K.A. Dill, J. Am. Chem. Soc. 120, 3166
(1998).

\bibitem{Tr99}T.M.Truskett, P.G. Debenedetti, S. Sastry and S. Torquato, J. Chem. Phys. 111,
2647 (1999).

\bibitem{Tr02}T.M.Truskett and K.A. Dill, J. Chem. Phys. 117, 5101 (2002).


\bibitem{Ku05}P. Kumar, S. V. Buldyrev, F. Sciortino, E. Zaccarelli and
H. E. Stanley, Phys. Rev. E {\bf 72}, 021501 (2005).


\bibitem{Fr03}G. Franzese, M.I. Marques and H.E. Stanley, Phys Rev E 67, 011103 (2003).

\bibitem{Ja98} E. A. Jagla, Phys. Rev. E {\bf 58}, 1478 (1998);
E. A. Jagla, J. Chem. Phys. {\bf 110}, 451 (1999); E. A. Jagla,
J. Chem. Phys. {\bf 111}, 8980 (1999); E. A. Jagla, Phys.
Rev. E {\bf
63}, 061501 (2001); E. A. Jagla, Phys. Rev. E {\bf 63}, 061509
(2001).

\bibitem{Wi02}N. B. Wilding and J. E. Magee, Phys. Rev. E 66, 031509 (2002). 

\bibitem{Ca03}P. J. Camp, Phys. Rev. E. {\bf 68}, 061506 (2003).

\bibitem{Ca05}P. J. Camp, Phys. Rev. E. {\bf 71}, 031507 (2005).
\end{thebibliography}
\end{document}